\def\year{2021}\relax
\documentclass[letterpaper]{article} % DO NOT CHANGE THIS
\usepackage{aaai21}  % DO NOT CHANGE THIS
\usepackage{times}  % DO NOT CHANGE THIS
\usepackage{helvet} % DO NOT CHANGE THIS
\usepackage{courier}  % DO NOT CHANGE THIS
\usepackage[hyphens]{url}  % DO NOT CHANGE THIS
\usepackage{graphicx} % DO NOT CHANGE THIS
\usepackage{amsfonts}
\usepackage{natbib}  % DO NOT CHANGE THIS AND DO NOT ADD ANY OPTIONS TO IT
\usepackage{caption} % DO NOT CHANGE THIS AND DO NOT ADD ANY OPTIONS TO IT
\usepackage{algorithm}
\usepackage{algpseudocode}
\title{Adversarial Attacks for Tabular Data: Application to Fraud Detection and Imbalanced Data}

\author {
	Francesco Cartella\textsuperscript{\rm 1 \rm *},
	Orlando Anuncia\c{c}\~{a}o \textsuperscript{\rm 1 \rm *},
	Yuki Funabiki \textsuperscript{\rm 1 \rm *}, 
	Daisuke Yamaguchi \textsuperscript{\rm 1}, \\
	Toru Akishita \textsuperscript{\rm 1}, 
	Olivier Elshocht \textsuperscript{\rm 1} \\
}
\affiliations {
	\textsuperscript{\rm 1} Sony Corporation \\
	\textsuperscript{\rm *} Joint first author \\
	Francesco.Cartella@sony.com, Orlando.Anunciacao@sony.com, Yuki.Funabiki@sony.com, Daisuke.C.Yamaguchi@sony.com, Toru.Akishita@sony.com, Olivier.Elshocht@sony.com
}

\begin{document}

\maketitle

\begin{abstract}
Guaranteeing the security of transactional systems is a crucial priority of all institutions that process transactions, in order to protect their businesses against cyberattacks and fraudulent attempts. Adversarial attacks are novel techniques that, other than being proven to be effective to fool image classification models, can also be applied to tabular data. Adversarial attacks aim at producing adversarial examples, in other words, slightly modified inputs that induce the Artificial Intelligence (AI) system to return incorrect outputs that are advantageous for the attacker. In this paper we illustrate a novel approach to modify and adapt state-of-the-art algorithms to imbalanced tabular data, in the context of fraud detection. Experimental results show that the proposed modifications lead to a perfect attack success rate, obtaining adversarial examples that are also less perceptible when analyzed by humans. %Results obtained on a real-world production system demonstrate also that the proposed techniques represent a serious threat to the robustness of advanced AI-based fraud detection procedures.
Moreover, when applied to a real-world production system, the proposed techniques shows the possibility of posing a serious threat to the robustness of advanced AI-based fraud detection procedures.

\end{abstract}

\section{Introduction}
\label{sec:Introduction}
Fraud detection plays a crucial role in financial transactional systems such as  banks, insurances or online purchases. The ability to detect early whether a transaction is fraudulent has a very high value and big investments have been made to make these systems more effective. It is however important to note that fraudsters are constantly developing new ways of fooling these systems, a phenomenon known as concept drift~\cite{widmer1996learning}. A fraud detection system therefore typically has high maintenance requirements. 

Machine Learning (ML) is a classical approach for fraud detection systems~\cite{abdallah2016fraud,ngai2011application}. The ability to retrain the models with new data helps in this need for adaptation to new fraud patterns. However, given the possibility of errors in the models decisions, which could lead to overlooking frauds or blocking licit transactions and sales opportunities, fraud detection systems often do not rely solely on the models but also contain one or more layers involving some form of human intervention~\cite{carcillo2018scarff,dal2017credit}.

%~\cite{carcillo2019combining,carcillo2018scarff,dal2017credit} %However, fraud detection systems often do not rely solely on the models but also contain one or more layers involving some form of human intervention~\cite{carcillo2019combining,carcillo2018scarff,dal2017credit}.
Risky transactions can be manually inspected and a decision is made whether those transactions should go through or should be blocked.

Fraudsters may use a wide range of techniques to bypass fraud detection systems. Among these techniques, adversarial attacks are novel and innovative approaches that might be used as a next level of smart financial frauds. The goal of adversarial attacks is to generate adversarial examples, i.e., inputs that are almost indistinguishable from natural data and yet classified incorrectly by the machine learning model~\cite{madry2017towards}. 

Algorithms to build adversarial examples have recently been shown to be very effective in fooling Machine Learning models, in particular Deep Neural Networks (DNNs) in Image Recognition~\cite{papernot2016limitations}. This is a cause of concern for many applications that rely on these technologies, such as self-driving cars or facial recognition. The reason adversarial examples exist is a consequence of the difference between the way humans and machines represent knowledge and relations of visual elements in object recognition tasks. This difference leads to the possibility for an attacker to perturb the pixels of an image in a way that the change is imperceptible to a human, but still induces an image classifier to produce a wrong interpretation. For instance, an attacker can induce an image classifier to recognize with very high confidence a gibbon in a picture that represents a panda, after the color of a few pixels has been slightly modified~\cite{goodfellow2015explaining}.

Recent studies~\cite{ballet2019imperceptible} have shown that adversarial algorithms can also be applied to other types of machine learning models using tabular data. The positive results obtained by these studies highlighted the need and the importance of investigating adversarial algorithms for a wider range of domains and applications, so that effective defensive strategies can be designed.

Motivated by the crucial role that security plays in the financial sector, in this paper we deal with the problem of creating adversarial examples for tabular data to effectively bypass fraud detection checks. In the particular case of this research, bypassing fraud checks means either inducing the system to classify a fraudulent transaction as non-fraud, or make the violation unnoticed by a potential human inspection. These kinds of misclassifications are particularly risky for fraud detection systems, as they would lead attackers to succeed in their criminal intent and to obtain illegal economic advantages. 

It is a well-known fact that the security of a system is related to the protection of multiple layers of an application~\cite{Zhu2011hierarchical}. Therefore, the security of a particular part of a system should be treated independently, without relying on the integrity of other layers. In light of this concept, in this paper we assume that the training set is available to the attacker, as our main focus is the analysis of security risks affecting the Machine Learning layer of a fraud detection system. Using this data, a proper surrogate model can be created and used to evaluate the effectiveness of the obtained adversarial examples, before submitting them to the real system. Notice that we do not make any assumptions on the architecture of the real model, which can be considered as unknown by the attacker.

To build successful attacks we tackled several problems, like adapting adversarial algorithms to imbalanced fraud detection data and properly treating non-editable variables. Moreover, as fraud detection systems often involve human intervention, we also considered the problem of building imperceptible adversarial examples, that are more difficult to be detected by operators. Experimental results show that, with the modifications introduced in this paper, it is possible to build realistic and imperceptible adversarial examples for all the fraudulent transactions of the considered use case. In comparison with state-of-the-art techniques, we achieved a drop of up to 64\% in the number of perturbed variables that are most commonly checked by human investigators and, for the most successful cases, adversarial examples were obtained by modifying just a minimum number of fields, reducing the probability for an attack to be discovered. Finally, by obtaining a 13.6\% success rate in attacking a deployed production system, we also demonstrated that the resulting adversarial examples were transferable to a target real-world model, representing a real threat to businesses dealing with fraud detection operations.

\section{Related Works}

Since the concept of adversarial examples was proposed~{\cite{szegedy2013intriguing}}, it has been a main topic in the field of adversarial machine learning.
This topic is especially discussed in image recognition tasks using DNNs, but has recently been discussed in other tasks such as audio recognition~{\cite{carlini2018audio}}, text summarization~{\cite{cheng2020seq2sick}}, neural network policy for reinforcement learning~{\cite{huang2017policies}} and so on.
Furthermore, the provision of tools to assist in the generation of adversarial examples is being promoted.
This is intended to make the training of models more robust against adversarial machine learning.
Adversarial examples can be easily generated using tools like Adversarial Robustness Toolbox~{\cite{art2018}}, CleverHans~{\cite{goodfellow2016cleverhans}}, Foolbox~{\cite{rauber2017foolbox}} and advertorch~{\cite{ding2019advertorch}}.
However, despite such trends, there are only a few studies of adversarial examples on tabular data.

To the best of our knowledge, the paper (a)~\cite{ballet2019imperceptible} was the first systematic introduction to adversarial examples in tabular domain using recent terminology. However, similar concepts, such as performing small changes in features to get desired outputs or executing model inversion, were studied before~\cite{grosse2017statistical,papernot2016transferability,bella2010data}.
%To the best of our knowledge, the concept of adversarial examples was first introduced to the tabular domain in the paper (a) by Ballet et al.~{\cite{ballet2019imperceptible}}.
Although it is very difficult to discuss the imperceptibility of perturbations in the tabular domain as opposed to the image domain, the authors of~\cite{ballet2019imperceptible} proposed to use the feature importance (the contribution of each feature to the model) as an indicator and applied more perturbations on the less important features.
Another recent paper (b)~{\cite{hashemi2020permuteattack}} treated the adversarial examples on tabular data as counterfactual examples to help the explainability of the model.
Another paper (c)~{\cite{levy2020notall}} proposed a method of conversion to a surrogate model that maintains the properties of the target model in order to apply existing generation methods.

The main differentiators of our research with respect to the contributions mentioned above consist of:
(a) they adopted a gradient-based attack method, which is applicable to models like DNNs but not to architectures with discrete gradients (such as decision trees), while we propose a model agnostic approach applicable to any architecture;
(b) they provide a counterfactual explanation for model interpretation, while we assume a more realistic scenario of attack attempts;
(c) the generation method they used is a variant of the black-box attack via surrogate model, while we assume that less information about the model is available to the attacker.

\section{Main Contributions}

As discussed in Section~\ref{sec:Introduction}, in this paper we present a novel approach to adapt adversarial attack algorithms, that are commonly used in the image recognition domain, to tabular data. In particular, we target fraud detection use cases. Achieving this goal requires facing and solving several challenges that arise from the different nature of data and model types used, compared to image classification applications.

One of the main differences between image and fraud detection data is the balance of samples that represent each class. While image classification is a multiclass problem where each of the classes is represented by a relatively similar amount of instances, fraud detection data sets are usually binary (i.e., they contain only two classes) and are typically characterized by a large imbalance between genuine and fraudulent transactions, the first being in big majority. Fraud detection models return a risk score that represents the estimated probability that the classified transaction is fraudulent. The large imbalance in the data leads to highly biased models that tend to attribute a risk score that takes into account the higher probability of observing instances belonging to the most numerous class. Differently from image classification, where the image is generally attributed to highest predicted probability class (with some exception, like in the case of diagnosis or fault detection applications), a decision threshold is normally tuned for fraud detection, according to some business requirements. The input transaction is deemed to be fraudulent if the predicted risk score is bigger than the threshold. In this research, we introduced the concept of the decision threshold within the attack algorithms, as it represents essential information to verify if the perturbations correctly resulted in the creation of adversarial samples.

Moreover, ensemble models and, in particular, Extreme Gradient Boosting techniques are commonly used in applications handling imbalanced tabular data, having proven to be particularly effective for these kind of settings. Applying adversarial attack algorithms in a model agnostic fashion, rather than closely targeted to Deep Neural Networks, was also one of the main challenges of this research.

Another aspect that differentiates image and tabular data is the value range that each feature can assume. Representing pixel values, image data can normally vary within limited ranges and data types (i.e., integer numbers between 0 and 255). On the other hand, tabular data can represent disparate pieces of information, like email addresses, surnames or amounts. As such, features representing a transaction can be extremely different from each other. Even if ultimately encoded as numerical values, a proper handling and representation of data types and range was essential to enable algorithms to generate realistic adversarial transactions.

Field editability also represents a crucial aspect to take into account when dealing with transactional data. Differently from the image domain, where an attacker can potentially modify any of the pixels independently, for tabular data there might be fields that are not directly controllable by the user but that are rather automatically determined by the system. Examples of these fields could be the historical amount borrowed in a loan application or the discount rate applied for an online purchase. To simulate the fact that direct changes to these values are not allowed by the system, specific constraints were added to the algorithm to prevent the modification of non-editable information.

Finally, we addressed imperceptibility as one of the main challenges of our research. Differently from image data, where imperceptibility is an intuitive concept related to human perception, for fraud detection we assume that imperceptibility is related to the number and entity of changes made to important features, such as fields that are most commonly checked by human operators within the specific application context. From a purely practical point of view, we define as imperceptible an adversarial attack that ultimately passes the fraud check, remaining unnoticed. We approached imperceptibility by introducing a custom norm as a measure of distance between the original transaction and the adversarial sample. The distance is obtained through weights assigned to each feature that are proportional to a novel definition of importance that takes into account the propensity of a feature to be inspected.  We will show that the custom norm properly drives the algorithm procedures to prioritize changes made on features that are rarely checked by human operators, obtaining less perceptible attacks.

Details of the contributions described above will be given in Section~\ref{sec:algomodification}, together with other aspects introduced in this research, including some algorithms specific solutions, like a novel loss function definition for the Zeroth Order Optimization (ZOO) algorithm~\cite{chen2017zoo} and an improved initialization strategy for Boundary attack~\cite{brendel2018decisionbased} and HopSkipJumpAttack~\cite{chen2020hopskipjumpattack}.

%including a compact representation of adversarial examples to visually evaluate attacks perceptibility and models vulnerabilities. Some algorithms specific solutions will be also discussed, like a novel loss function definition for the Zeroth Order Optimization algorithm~\cite{chen2017zoo} and an improved initialization strategy for Boundary attack~\cite{brendel2018decisionbased} and HopSkipJumpAttack~\cite{chen2020hopskipjumpattack}.

\section{Problem Statement}
\label{sec:problemstatement}

In this paper we address the problem of building adversarial examples for fraud detection systems on financial data.

A financial transaction is a vector of $m$ variables, $v_1,...,v_m$ with each $v_i \in \mathbb{R}$\footnote{Some variables can be textual, booleans, naturals or integers, but for the sake of simplicity we assume that there is a feature processing step that transforms all values into real numbers.}.

Fraud detection systems in financial data analyze a set of $n$ transactions $t_1,...,t_n$ in a certain period of time. A model $M$ is used to label each transaction $t_i$ with a corresponding class $c_i \in \{0, 1\} $ in which 0 corresponds to Non-Fraud and 1 corresponds to Fraud: $M(t_i)=c_i$.

The first goal of an attacker is to find a perturbation vector $ p_i=[ p_i^1,...,p_i^m ] $ such that:

\begin{equation}
M(t_i+p_i)=0
\end{equation}

for values of $i=1,...,n$ such that $M(t_i)=1$ and $t_i$ is a real fraud.

If such a perturbation vector $p_i$ can be found then $\tilde{t_i}=t_i+p_i$ is a successful adversarial sample.

For the purpose of our experiments, our goal is to create an adversarial sample $\tilde{t_i}$ such that $M(\tilde{t_i})=0$ for each fraudulent transaction $t_i$ that is correctly identified by model $M$ ($M(t_i)=1$).

\section{Algorithms modifications}
\label{sec:algomodification}

In this section we describe the main problems that were faced to create successful adversarial examples. We used the Adversarial Robustness Toolbox (ART)~\cite{art2018} as the reference tool. ART is a Python library for Machine Learning Security that provides tools to enable developers and researchers to defend and evaluate Machine Learning models and applications against adversarial attacks.

Even though some ART algorithms can be applied to tabular data, the majority of the tool's algorithms is designed to deal with image data. So it was no surprise that it was necessary to make changes in order to build successful adversarial examples for tabular data, and more specifically fraud detection.

\subsection{A Generic Adversarial Algorithm}

Adversarial algorithms can be used by attackers to retrieve the optimal changes that, when applied to the fraudulent transactions they want to submit, induce the fraud check to fail, by erroneously accepting the submitted transactions as legitimate. To simulate this scenario in our experimental setup, we applied adversarial algorithms on fraudulent samples that are correctly detected as fraud by the model under attack. An adversarial algorithm is considered successful if it outputs adversarial examples that are classified as non-frauds by the same model. A generic adversarial algorithm starts with an initial sample and makes perturbations to that sample until the model misclassifies it. A second goal of adversarial algorithms is to make the adversarial sample as similar as possible to the original sample.
%Adversarial algorithms get a fraudulent sample that is correctly detected as fraud by the model and output an adversarial sample that is wrongly classified as non-fraud by the same model. A generic algorithm starts with an initial sample that is drawn randomly and makes perturbations to that sample until the model misclassifies it. Adversarial algorithms have one main goal that is to build a sample that is adversarial. A second goal of adversarial algorithms is to make the adversarial sample as similar as possible to the original sample. 
Algorithm~\ref{alg:adversarialalgorithm} shows the pseudo-code of a simple generic algorithm that serves the purpose of illustrating the main concepts. This algorithm receives a fraudulent transaction $t$, a model $M$ such that $M(t)=1$ and a similarity threshold $\rho$. The perturbations can be selected through many different ways. Two of the most common approaches involve the use of distance metrics to get the adversarial sample $\tilde{t}$ \emph{closer} to the original sample $t$ (e.g.: Boundary or HopSkipJump) or calculations based on the gradient of the model (e.g.: ZOO algorithm). Algorithm~\ref{alg:adversarialalgorithm} uses a similarity function and a similarity threshold $\rho$. The similarity function can be based on the distance between $t$ and $\tilde{t}$, calculated using a norm such as $L_2$ or $L_\infty$. Threshold $\rho$ can be provided explicitly, as is the case in our generic algorithm. However, some algorithms calculate it in an indirect way. As an example, Boundary attack algorithm converges when it is close enough to the decision boundary. Algorithms usually also have a maximum number of allowed steps in the while loop. This was not included in the pseudo-code for simplicity reasons.

\begin{algorithm}
\caption{Generic Adversarial Algorithm}
\begin{algorithmic}[1]
\Function{generate\_adv}{$t,M,\rho$}
\State $\tilde{t} \leftarrow initialize\_sample()$
\While{$M(\tilde{t})=1 \lor similarity(\tilde{t},t) < \rho$}
\State     $\tilde{t} = make\_perturbation(\tilde{t},t)$
\EndWhile

\Return $\tilde{t}$
\EndFunction

\end{algorithmic}
\label{alg:adversarialalgorithm}
\end{algorithm}

\subsection{Using Custom Threshold}
\label{sec:threshold}

In ART, adversarial algorithms are fed with the model that is being attacked. Adversarial examples are iteratively refined and, at every iteration, the model is used to evaluate the current samples' success. Because fraud detection is binary, the adversarial algorithms stop as soon as the current adversarial example is deemed to be successful, i.e., when it is classified as non-fraud with a score higher than $0.5$. While this may work well for an image recognition model, it is problematic for a fraud detection model.

In fraud detection use cases, a decision threshold $\tau \in [0,1]$ is commonly tuned and an input transaction is classified as fraud if $[M(t)]_1 > \tau $,where $[M(t)]_1$ is the probability that the transaction $t$ belongs to class 1, i.e., fraud. The threshold $\tau$ is typically very small and much lower than 0.5, to compensate for the tendency of the model to attribute very low risk scores to new transactions, given the big majority of non-fraud samples observed at training time.

Our initial results when applying the default ART algorithms were poor because a threshold of 0.5 was used, misleading the algorithms by assuming that a successful adversarial sample had been found. To correct the problem we modified the Boundary~\cite{brendel2018decisionbased}, HopSkipJumpAttack~\cite{chen2020hopskipjumpattack} and ZOO~\cite{chen2017zoo} attacks. These algorithms are now fed with a custom threshold and whenever the model is evaluated internally, the custom threshold is taken into account for the models' decision. With this correction, the adversarial algorithms have access to true information about whether a sample is classified as fraud or not by the model.

\subsection{Specifying a Custom Loss Function for ZOO}

To drive the creation of adversarial examples, the ZOO algorithm uses a specific loss function that, as detailed below, implicitly considers a balanced threshold of 0.5 in its standard formulation. For this reason, the introduction of a novel loss function was essential to adapt the algorithm to biased cases.
%Specifying a threshold was not sufficient to make ZOO algorithm achieve very high success rate in our experiments. The reason for this is that ZOO algorithm uses a specific loss function that drives the creation of adversarial samples. Modifying the loss function was also essential to adapt the algorithm to biased cases.

%Formulated as a black box attack, the ZOO algorithm is independent from the target model internals and it can be applied to any model architecture, as soon as the output score is available.
To adapt the ZOO algorithm formulation to the specific case of binary classification and fraud detection, following the notation introduced in Section~\ref{sec:problemstatement}, let us define the model under attack as a function $M(t)$ that takes a transaction $t$ and returns a two dimensional vector $M(t) \in [0,1]^2$. The two dimensions of this vector represent the probability score of class 0 (not fraud) and of class 1 (fraud), respectively. As a consequence, $[M(t)]_0 + [M(t)]_1 = 1$

Given a fraudulent transaction $t_f$ correctly classified by the model, the ZOO attack finds the corresponding adversarial sample $\tilde{t}$ by solving the following optimization problem:
\begin{equation}
	\label{eq:zoomin}
    \mbox{minimize}_{\tilde{t}} \left[||\tilde{t} - t_f||_2^2 + r \cdot f(\tilde{t}) \right]
\end{equation}
where $||v||_2 = \sqrt{\sum_{i=1}^m v_i^2}$ denotes the Euclidean norm (or the $L_2$ norm) of the vector $v=[v_1,...,v_m]^T$ and $r > 0$  is a regularization parameter. Equation~\ref{eq:zoomin} is expressed as a sum of two terms to be minimized: the first term $||\tilde{t} - t_f||_2^2$ represents a measure of distance between the adversarial example $\tilde{t}$ and the original transaction $t_f$; the $f(\tilde{t})$ of the second term represents a loss function that measures how unsuccessful an adversarial attack is. The minimization of Equation~\ref{eq:zoomin} is performed using stochastic coordinate descent methods (see~\cite{chen2017zoo} for details).
The loss function proposed in the standard formulation of the ZOO algorithm is the following:
\begin{equation}
	\label{eq:zoooriglossfunction}
	f(t) = \mbox{max}\left[(log[M(t)]_1 - log[M(t)]_0),-\nu\right]
\end{equation}
where $\nu >=0$ is a tuning parameter for attack transferability, commonly set to 0 for attacking a targeted model or to a larger value when performing a transfer attack. If, for simplicity, we consider  $\nu=0$, the loss function above will return its minimum value of 0 for all the adversarial samples $\tilde{t}$ having $[M(\tilde{t})]_0 >= [M(\tilde{t})]_1$, i.e., probability of \textit{not fraud} bigger or equal than \textit{fraud}.
As explained previously, in the context of biased models, assigning to a transaction a not fraud probability higher than the probability of fraud, does not necessarily imply that the transaction is classified as licit, but it is necessary that $[M(t)]_1 \leq \tau$, where $\tau \in [0,1]$ is the decision threshold. As a consequence, the loss function of Equation~\ref{eq:zoooriglossfunction} is minimized also by a set of adversarial examples that, being still classified as fraud, are unsuccessful. This is the set of adversarial examples $\tilde{t}$ for which $[M(\tilde{t})]_0 \geq [M(\tilde{t})]_1$ and $[M(t)]_1 > \tau$.

As Equation~\ref{eq:zoooriglossfunction} results inadequate for imbalanced use cases, we propose to use the following loss function in the optimization of Equation~\ref{eq:zoomin}:
\begin{equation}
\label{eq:zoonewloss}
f(t) = \mbox{max} \left[\left([M(t)]_1 - \tau \right), -\nu\right]
\end{equation}
The loss function above assures that minimum values are obtained only for successful adversarial examples $\tilde{t}$, for which $[M(\tilde{t})]_1 <= \tau$ (i.e., classified as not frauds).
%It is important to notice that the parameter $\nu$ is still present in Equation \ref{eq:zoonewloss}, to explicitly impose a stronger constraint that would increase the chances of success in the case of a transfer attack.

\subsection{Creating Realistic Attacks with Editability Constraints}
\label{sec:realistic}

We analyzed the nature of the perturbations that were obtained by the adversarial algorithms. In particular, tabular data has features of different types: boolean, integer, hot-encoded variables, integers that only take positive values, etc. Without imposing any constraint, the adversarial algorithms created perturbations that led to illegal values, with respect to the type of features that are taken into consideration (e.g.: a boolean feature having value different from $0$ or $1$, or a positive integer feature that becomes negative). It was then necessary to make sure that perturbations assume only what we designated by \emph{realistic} values. Each variable $v_i$ can assume values from a specific domain $D_i$ (e.g.: for a real variable $v_i$, $D_i=\mathbb{R}$). An adversarial sample $\tilde{t}$ has a realistic value $x$ for variable $v_i\in D_i$ if $x \in D_i$. In the case that $x \notin D_i$ a transformation needs to be made in order to ensure that $x \in D_i$.

The inspection of adversarial samples raised awareness about the presence of non-editable fields in the data (i.e., fields that cannot be directly modified by the user), but are rather calculated automatically by the system. An example of this could be the total amount of money borrowed by a customer in the past, in the context of a loan management application. This value cannot be changed when a new loan is requested. Adversarial algorithms should take this into account and only make changes to variables that the user can have access to. In order to address this we defined an editability vector that contains the variables that can be changed by adversarial algorithms.

In order to address realistic and editability problems, we modified the adversarial algorithms ZOO, Boundary and HopSkipJump. In the execution of each algorithm, whenever a potential adversarial sample is modified, editability and realistic properties are enforced by correcting the illegal values.

In order to make adversarial samples realistic we considered the data types and the corresponding corrections for a specific value $x$ that are listed on Table~\ref{tab:makerealistic}:

\renewcommand{\arraystretch}{1.2}

\begin{table}[h!]
\centering
\begin{tabular}{|c|c|}
\hline
\textbf{Type} & \textbf{Correction}\\
\hline
Boolean & 0 if $x \leq 0.5$, 1 otherwise \\
\hline
Integer & $\mathrm{round}(x)$ \\
\hline
Positive Integer & $\mathrm{round}(x)$ if $x \geq 0$, 0 otherwise\\
\hline
Positive Float & 0 if $x < 0$, $x$ otherwise \\
\hline
Hot-encoded fields & 1 for field with maximum value.\\
 & 0 for other fields of same group \\
\hline 
\end{tabular}
\caption{Data types and corresponding corrections for adversarial samples}
\label{tab:makerealistic}
\end{table}

In order to implement corrections listed on Table~\ref {tab:makerealistic}, adversarial algorithms now receive a data specification dictionary containing a list of features for each data type.

The editability constraints are enforced by defining a vector of editable features and passing it to the adversarial algorithms. The editability vector $e$ for variables $v_i,...,v_m$ is defined as $e_i=1$ if $v_i$ is editable, $0$ otherwise, for $i=1,...,m$. Algorithms will only allow perturbations on features $v_i$, with $i=1,...,m$ for which $e_i=1$. Features $v_j$, with $j=1,...,m$ for which $e_j=0$ are not perturbed and forced to maintain their original values. Which feature $v_i$ are editable is a property of the system under consideration.

\subsection{Specifying a Custom Norm}
\label{sec:customnorm}

After creating realistic adversarial samples and taking editability into consideration, it was important to go one step further in terms of imperceptibility of the attack. Besides editability considerations, adversarial algorithms pick up any available feature as a candidate for a perturbation. Within a specific application context, an attacker can guess that, in the case of a hypothetical manual inspection, some features may capture the attention of human operators more than others. For instance, in a loan request application, the applicant salary information is usually more informative than other fields, like the number of owned pets~\cite{ballet2019imperceptible}. Nevertheless, less important features are also considered by the model to estimate the request's risk score. As a consequence, the attacker's goal is to minimize the perturbations made on features that have a bigger chance to be checked. 

Adversarial algorithms such as Boundary or HopSkipJump attacks use norms as measures of distance between adversarial and original examples. These algorithms try to minimize this distance as much as possible in order to make the adversarial example imperceptible. $L_2$ norm considers the global distance between the original and the adversarial sample, disregarding that some features may have very large perturbations. Minimizing $L_\infty$ on the other hand means that the algorithm will try to avoid having a big perturbation on a single feature, giving preference to small perturbations on many features. None of these norms completely satisfy the needs of an imperceptible attack in the context of tabular data. In order to do that more successfully it is necessary to consider features differently, depending on whether they are checked by a human operator. This motivated the introduction of a novel custom norm that is expressed in Equation~\ref{eq:customnorm}:

\begin{equation}
\label{eq:customnorm}
n=||p(\alpha h+\beta[(1-h)(1-v)+hv])||_\gamma
\end{equation}

where $p$ is the perturbation vector, $h$ is a Boolean vector indicating whether a variable is checked, $v$ is a vector of feature importance, $\alpha,\beta \in [0,1]$ are weights on the check and importance of a feature respectively, and $||.||_\gamma$ is a $\gamma$-norm such as $L_2$ norm that is being used. It is known that algorithms using gradient descent can empirically derive values of coefficients such as $\alpha$ and $\beta$ in a binary search~\cite{carlini2017robustness}, but it is future work to verify whether these techniques are applicable to our approach.

For the definition of the custom norm, two properties were considered: 1) whether a feature is checked or ignored by the operators and 2) the importance of the feature for the model. The idea behind the custom norm is that changes to features that are checked and important lead to high values of the distance, so that the optimization algorithm prefers other solutions. Moreover, we also want to penalize solutions in which the feature is not checked and not important, because it will not have a significant effect in the attack. On the other hand, we would like the algorithm to prefer solutions based on perturbing features that are not checked and have high importance for the model. For these types of perturbations the custom norm returns low values. Finally, if checked variables need to be perturbed it is preferable that they are not important for the model, so we assign low distances for these situations. In conclusion, the goal of the custom norm is to drive the optimization procedure of the attack algorithms to obtain adversarial examples that are imperceptible and unnoticed by human operators.
%The goal of the custom norm is to drive the optimization procedure of the attack algorithms to perturb variables in a way that the obtained adversarial example results imperceptible and unnoticed by human operators. If checked variables need to be perturbed it is preferable to do it to not important variables, so our custom norm will give a low value in these situations.

\section{Experiments and Results}
\label{sec:expresults}

In this section we describe the experiments that were performed and the obtained results.
After having modified the ART algorithms as discussed in Section~\ref{sec:algomodification}, we applied them to the \textit{German Credit Dataset}~\cite{dua2017gcd} use case. The strategy described in the following sections was also applied to 2 additional datasets with similar results. The results obtained using the first dataset (the IEEE-CIS Fraud Detection dataset) are not detailed due to space limitations. The second dataset is an internal dataset that cannot be disclosed for confidentiality reasons.

\subsection{Use Case and Data Preparation}

\textit{German Credit Dataset}~\cite{dua2017gcd} is a publicly available dataset used for building models that evaluate the risk of a loan application, given account and customer information.
Out of 1000 applications in total, 700 were accepted while 300 were rejected and deemed to be risky in terms of low propensity of the applicant of being able to pay back the loan. In the context of adversarial attacks, we considered the rejected applications as fraudulent, as the goal of a potential attacker would be to slightly modify their loan request such that it eventually gets accepted.
The data contains 20 features with 7 integer and 13 categorical ones, such as age, sex, purpose of the loan or if the customer is a foreign worker.
We applied a one-hot encoding to categorical features, obtaining a total of 61 numerical features for modeling.

\paragraph{Capabilities of Attackers}

As discussed, the goal of attackers is to modify true-positive requests (i.e., applications that are deemed risky and should not be accepted) so that they can be accepted.
Our assumption is that the attacker can make reasonable judgments about the importance of the features and estimate what are the fields that a human investigator most probably checks to measure the applicant's ability to pay back the loan. In this experiment, we assumed that human investigators would mainly check 10 of the total 20 features such as the \textit{``Purpose (of the loan)"} and \textit{``Credit amount"}.
Moreover, we assumed that the features \textit{``Credit history"}, \textit{``Personal status and sex"}, \textit{``Other debtors/guarantors"} and \textit{``Age in years"} are not directly modifiable by the attacker and set them as non-editable.
%Moreover, we assumed 4 non-editable features, \textit{``Credit history"}, \textit{``Personal status and sex"}, \textit{``Other debtors/guarantors"} and \textit{``Age in years"} as the information that the attacker was unable to directly modify.
Although we conducted experiments under these hypothesis, different settings can be considered as well, depending on different assumptions on the system and the application context.
%In this paper, we conducted experiments with theses assumptions, but we can apply the proposed attacks under different assumptions.

\paragraph{Model Construction}

We used XGBoost~\cite{Chen2016xgboost} as a learning algorithm.
At first, the dataset was split into train and test sets consisting of 70\% and 30\% of the data, respectively.
Furthermore, train set was split into training and validation sets consisting 80\% and 20\% of the data, respectively.
The training data was used to generate a binary classification model and the validation data was used to adjust the threshold.
Before the threshold adjustment, the accuracy on validation set was 75.7\%, the recall was 38.1\% and the precision was 66.6\%. An optimal threshold of $0.192$ was obtained using the F2 score maximization as a target metric. With this threshold we obtained an accuracy of 60.0\%, a recall of 95.2\% and a precision of 42.6\%.
%We used the F2 score to calculate the threshold and obtained 0.19217326 as the optimal threshold.
%As a result, the accuracy decreased to 60.0\%, but the recall increased to 95.2\%.
Using the resulting model on the test set, we were able to discriminate 82 true-positive data, representing a recall of 91\% and a precision of 42.5\%. These results show that, even without performing particularly sophisticated feature engineering, we obtained a fair model with satisfactory performance that can be effectively used to evaluate our study.

\subsection{Results}

%In this subsection we summarize the results obtained. Our goal is to show that the approaches to the problem we faced in building adversarial samples for tabular data in the fraud detection domain were successful.
In this subsection we summarize the results obtained. Our goal is to show that the approaches we followed to build adversarial samples were successful.

We considered 4 parameters that can be switched on and off in our experiment designs: threshold (Section~\ref{sec:threshold}), realistic (Section~\ref{sec:realistic}), editability (Section~\ref{sec:realistic}) and custom norm (Section~\ref{sec:customnorm}). 
%We consider 5 different configurations using these parameters as shown in Table~\ref{tab:experimentconfigurations}.
As shown in Table~\ref{tab:experimentconfigurations}, we obtained 5 different configurations.

\begin{table}[h!]
\centering
\begin{tabular}{|c|c|c|c|c|}
\hline
\textbf{ID} & \textbf{Threshold} & \textbf{Realistic} & \textbf{Custom} & \textbf{Editability}\\
& & & \textbf{Norm} &\\
\hline
1 & OFF & OFF & OFF & OFF\\
\hline
2 & ON & OFF & OFF & OFF\\
\hline
3 & ON & ON & OFF & OFF\\
\hline
4 & ON & ON & ON & OFF\\
\hline
5 & ON & ON & ON & ON\\
\hline
\end{tabular}
\caption{Configurations of the performed experiments}
\label{tab:experimentconfigurations}
\end{table}

Experiment 1 is performed with ART as it is, without any changes or adaptations. In Experiment 2 we use custom thresholds as described in Section~\ref{sec:threshold} and in Experiment 3 we make the attacks realistic (Section~\ref{sec:realistic}). In Experiment 4 we use the custom norm as described in Section~\ref{sec:customnorm} and in Experiment 5 we add editability constraints (Section~\ref{sec:realistic}).

The results obtained are shown in Tables~\ref{tab:resultsboundary},~\ref{tab:resultshsj} and ~\ref{tab:resultszoo}.

\begin{table}[h!]
\centering
\begin{tabular}{|c|c|c|c|c|c|}
\hline
\textbf{Boundary} & \textbf{1} & \textbf{2} & \textbf{3} & \textbf{4} & \textbf{5}\\
\hline
Success Rate (\%) & 0 & 100 & 100 & 100 & 100\\
\hline
\% of Unrealistic Values & - & 47.6 & 0 & 0 & 0 \\
\hline
\# Checked Fields & - & - & 592 & 214 & 228\\
\hline
\# Non-Editable Fields & - & - & - & 63 & 0 \\
\hline
\end{tabular}
\caption{Results obtained with each experiment configuration for Boundary attack}
\label{tab:resultsboundary}
\end{table}

\begin{table}[h!]
\centering
\begin{tabular}{|c|c|c|c|c|c|}
\hline
\textbf{HopSkipJump} & \textbf{1} & \textbf{2} & \textbf{3} & \textbf{4} & \textbf{5}\\
\hline
Success Rate (\%) & 0 & 100 & 100 & 100 & 100\\
\hline
\% of Unrealistic Values & - & 87.0 & 0 & 0 & 0\\
\hline
\# Checked Fields & - & - & 554 & 465 & 418\\
\hline
\# Non-Editable Fields & - & - & - & 159 & 0 \\
\hline
\end{tabular}
\caption{Results obtained with each experiment configuration for HopSkipJump attack}
\label{tab:resultshsj}
\end{table}

\begin{table}[h!]
\centering
\begin{tabular}{|c|c|c|c|c|c|}
\hline
\textbf{ZOO} & \textbf{1} & \textbf{2} & \textbf{3} & \textbf{4} & \textbf{5}\\
\hline
Success Rate (\%) & 0 & 100 & 100 & 100 & 100\\
\hline
\% of Unrealistic Values & - & 1.6 & 0 & 0 & 0\\
\hline
\# Checked Fields & - & - & 159 & 133 & 153\\
\hline
\# Non-Editable Fields & - & - & - & 37 & 0 \\
\hline
\end{tabular}
\caption{Results obtained with each experiment configuration for ZOO attack}
\label{tab:resultszoo}
\end{table}

Experiment 1 was very unsuccessful, with no adversarial samples found. This means that the original algorithms cannot be applied directly to an unbalanced problem. When changes are made to the algorithm to use a proper threshold in Experiment 2, the success rate increases to 100\% for the three algorithms. This experiment however still generates unrealistic values for some features. As an example, Table~\ref{tab:resultshsj} shows that 87\% of the values generated by HopSkipJump are unrealistic. This makes it easy for a human operator to detect the attack. In Experiment 3 this problem is solved and only realistic values are generated. Experiment 3 does not use the custom norm, which is done in Experiment 4. By observing the results we can check that, by using the custom norm, the number of perturbed checked fields decreased in the application of each of the 3 adversarial algorithms, thus increasing imperceptibility. For instance, Table~\ref{tab:resultsboundary} shows that, in Experiment 3, a total of 592 fields that are checked by human operators were perturbed by the boundary attack. When the proposed custom norm is used in Experiment 4, only 214 of these fields are modified, representing a drop of 64\%, with respect to the state-of-the-art norm. Experiment 5 considers editability constraints and we observe that the number of non-editable fields that are changed is reduced to 0 in each algorithm. For Boundary and ZOO attacks on Experiment 5 there is a slight increase on the number of checked fields that are changed. This can be explained by the fact that the algorithms are not allowed to change non-editable fields and the pressure to change checked fields is higher.

Finally, it is important to mention that some successful adversarial examples were obtained by changing just a few fields. Table~\ref{tab:adversarialsample} shows an example where changing only the value of one attribute (\textit{Status checking account}) caused the model to return a lower risk score and flip its decision from rejection to the acceptance of the loan application. It is evident that these types of adversarial examples are highly imperceptible and that it is very probable that they might remain unnoticed.

\begin{table}[h!]
	\centering
	\begin{tabular}{|c|c|c|}
		\hline
		\textbf{ZOO algorithm} & \textbf{Original} & \textbf{Adversarial}\\
		\hline
		Status checking account & A12 & A14 \\
		\hline
		Model's Risk Score & 0.275 & 0.127 \\
		\hline
	\end{tabular}
	\caption{One adversarial example obtained with ZOO algorithm.}
	\label{tab:adversarialsample}
\end{table}

\section{Conclusions and Future Work}
In this paper we illustrated the process we followed to adapt state-of-the-art adversarial algorithms, that are commonly used in the image classification domain, to imbalanced tabular data. In particular we targeted fraud detection use cases.

After verifying the inadequacy of existing techniques to handle tabular data, we introduced modifications to address the shortcomings. In particular (i) we allowed adversarial algorithms to deal with biased model scores through the usage of a custom threshold within the algorithms and the introduction of a novel loss function for ZOO algorithm; (ii) we introduced constraints in the allowed perturbation to obtain realistic adversarial examples and avoid out-of-bound values; (iii) we improved imperceptibility through a proper management of not editable fields and through the introduction of a custom norm that drives the creation of adversarial examples that have a higher chance to be unnoticed by human investigators.

In terms of results, the changes we made contributed to increase the attack success rate from 0\% to 100\%. Moreover we showed examples of successful imperceptible attacks that were obtained by changing the value of just a few features.

Ultimately, we conducted a final experiment on the transferability of the adversarial examples to a real-world production system. To this extent, we could not perform attack transferability for the use case we considered in this paper, given the lack of a real deployed AI system. For this reason, we executed the full adversarial attack process on a real-world use case that is currently in production. For confidentiality reasons and due to the substantial economical dangers that sharing information on internal system vulnerabilities might cause, only final results can be reported, without disclosing any detail about the analyzed use case. We submitted 44 modified fraudulent transactions, created using a surrogate side model, to the real production system. For 35 transactions, representing 80\% of the submitted adversarial examples, the production model returned a lower risk score than for the original transaction. More importantly, 6 cases, representing 13.6\% of the submitted transactions, were flagged as safe by the system and automatically accepted, bypassing the human check. These results demonstrate that the techniques introduced in this paper represent a real threat for many AI-based fraud detection models, used in day-to-day business.

Future work will be conducted in the direction of performing more extensive experiments on attack transferability, by setting a lower target threshold for the adversarial algorithms, in order to increase the success probability of attacks for the considered real-world use case. 

On the other hand, these preliminary results highlighted the need of assuring a better robustness of production fraud detection models. To this extent, we started exploring the topic of defense techniques, with the goal of improving their ability to detect and block also the most sophisticated adversarial attacks. After conducting a survey of existing defensive methodologies, our plan is to identify their shortcomings and, eventually, come up with new approaches, following a similar process we used for the attack techniques. 
%The ultimate goal of this research is to increase the security and robustness of AI-based production systems against adversarial attacks. 

% References and End of Paper
% These lines must be placed at the end of your paper
\fontsize{9.0pt}{10.0pt} \selectfont
\bibliography{paper}
\end{document}